\documentclass[aps,pra,preprint,showpacs,amsmath,amssymb,groupedaddress,floatfix,nofootinbib] {revtex4-1}

\usepackage{graphicx}
\usepackage{dcolumn}
\usepackage{bm}
\usepackage{amssymb}
\usepackage{multirow}
\usepackage{SIunits}
\usepackage{float}
\usepackage{color} 

\begin{document}

\title{Beyond mean-field dynamics of ultra-cold bosonic atoms in higher dimensions: facing the challenges with a multi-configurational approach }

\author{V. J. Bolsinger} \email{vbolsing@physnet.uni-hamburg.de}
\author{S. Kr\"onke}
\author{P. Schmelcher}

\affiliation{Zentrum f\"ur Optische Quantentechnologien, Universit\"at Hamburg, Luruper Chaussee 149, 22761 Hamburg, Germany}
\affiliation{The Hamburg Centre for Ultrafast Imaging, Universit\"at Hamburg, Luruper Chaussee 149, 22761 Hamburg, Germany}

\date{\today}

\pacs{pacs code}

\begin{abstract} 

Exploring the impact of dimensionality on the quantum dynamics of interacting bosons in traps including particle correlations is an interesting but challenging task. Due to the different participating length scales the modelling of the short-range interactions in three dimensions plays a special role. We review different approaches for the latter and elaborate that for multi-configurational computational strategies finite range potentials are adequate resulting in the need of large grids to resolve the relevant length scales.
This results in computational challenges which include also the exponential scaling of complexity with the number of atoms. We show that the recently developed ab-initio Multi-Layer Multi-Configurational Time- Dependent Hartee method for Bosons (ML-MCTDHB) [J. Chem. Phys. 139, 134103 (2013)] can face both numerical challenges and present an efficient numerical implementation of ML-MCTDHB in three spatial dimensions, particularly suited to describe the quantum dynamics for elongated traps.

The beneficial scaling of our approach is demonstrated by studying the tunnelling dynamics of bosonic ensembles in a double well.
Comparing three-dimensional with quasi-one dimensional simulations, we find dimensionality-induced effects in the density.
Furthermore, we study the crossover from weak transversal confinement, where a mean-field description of the system is sufficient, towards tight transversal confinement, where particle correlations and beyond mean-field effects are pronounced. 

\end{abstract}

\maketitle


\section{Introduction}

Trapped ultracold bosonic gases constitute intriguing  quantum many-body systems, well-known for their tunability \cite{Pitaevskii2003,Pethick2008}.
They allow for mimicking the physics of e.g. a diversity of condensed matter problems \cite{Bloch2008b} as well as for addressing fundamental questions.
One of the latter is how dimensionality influences the dynamics of interacting many-body systems.
Many parameters can be tuned externally, for example, the interaction strength via Feshbach resonances \cite{Chin2010},  making a transition form weakly interacting systems to unitarity possible \cite{Braaten2006,Astrakharchik2004}, or the dimensionality by individually adapting the trapping strengths in the different spatial directions \cite{Reichel2011,Metcalf1999,Dettmer2001,Gorlitz2001, Moritz2003,Henderson2009}, allowing for the study of the cross over from 3D to quasi 1D.

For this presented purpose, we regard a system as quasi one-dimensional, if the transversal degrees-of-freedom can be adiabatically separated\footnote{With adiabatic separation, we mean a crude adiabatic separation of the transversal coordinates $\rho_i=(x_i,y_i)^T$ throughout this work, i.e.\ $\Psi(\mathbf{r}_1,...,\mathbf{r}_N)=\phi_{\parallel}(z_1,...,z_N) \phi_{\bot}(\rho_1,...,\rho_N)$. We remark that the term ``quasi one-dimensional'' is often used in a stricter sense, namely if a closed theory for the longitudinal degrees-of-freedom $\phi_{\parallel}(z_1,...,z_N)$ can be formulated.}, 
which is the case in elongated traps if the energy of the transversal first excited state is much higher than all other relevant energy scales such that it cannot be populated.
Then, there are no spatial correlations between the longitudinal and transversal directions.
However strong transversal confinement can still influence the quasi one-dimensional system, for instance, by confinement induced resonances \cite{Olshanii1998,Bergeman2003}, where transversal  modes are virtually excited, or by geometric potentials in the case of curved wave guides \cite{DaCosta1981,Reitz2012,Stockhofe2014,Exner2015}.
Regarding many-body effects, there are intriguing differences between quasi one-dimensional and three-dimensional systems, which strongly motivates research on the crossover regime:
By dimensional arguments in the thermodynamic limit for uniform systems, one can show that the ratio of interaction to kinetic energy scales like $n_{3D}^{1/3}g$ in three dimensions with $n_{3D}$ denoting the particle density and $g$ interaction strength being proportional to the s-wave scattering length, while in one dimension the scaling goes as the so-called Lieb-Liniger parameter $g/n_{1D}$ with the one-dimensional particle density $n_{1D}$ \cite{Lieb1963}.

Three-dimensional, weakly interacting bosonic systems can be well described by the Gross-Pitaevskii mean-field equation (GPE) \cite{Gross1961,Pitaevskii1961}, to study e.g. collective excitations (see \cite{Pitaevskii2003,Pethick2008} and refs. therein) or vortices \cite{Kevrekidis2008}.
In one-dimensional weakly interacting systems, other types of stable mean-field excitation can arise, such as dark and bright matter-wave solitons (see \cite{Kevrekidis2008} and refs. therein).

Thereby, the cross over from one to three dimensions is of particular theoretical and experimental interest: In this regime, one may study e.g.\ the decay of quasi one-dimensional dark soliton excitations into three-dimensional entities such as vortex rings or solitonic vortices (see \cite{Donadello2014} and refs. therein)
or the transition from a 1D quasi-condensate to a 3D condensate \cite{Armijo2011,Yang2014}.

If the Lieb-Liniger parameter is increased to $g/n_{1D} \gtrsim 1$ or enough excitation energy is provided, one-dimensional quantum gases enter a completely different regime where the (quasi-) condensate description and quasi-1D GPE approach breaks down and intriguing correlation effects emerge, for instance, the fermionization of strongly interacting bosons \cite{Giradeau1960,Blume2002}, fragmentation \cite{Mueller2006} or the decay of dark solitons due to dynamical quantum depletion (\cite{Delande2014,Kronke2015} and refs. therein).
Recently, a strong interest in unravelling correlation effects also in two- and three-dimensional systems has emerged involving the study of the relation between mean-field vortices and the exact solution of the linear many-body Schr\"odinger equation \cite{Dagnino2009,Cremon2015}, beyond mean-field effects in vortices \cite{Klaiman2014,Weiner2014,Klaiman2016,Tsatsos2015}, as well as other fragmentation scenarios \cite{Fischer2015,Streltsova2014a,Sakmann2014,Streltsov2013}.

In this work, we show how the recently developed  Multi-Layer Multi-Configurational Time-Dependent Hartree method for Bosons (ML-MCTDHB) \cite{Cao2013, Kronke2013} can be applied to efficiently study the correlated quantum dynamics of short-range interacting ultracold bosonic atoms in three-dimensional elongated traps. For such simulations, two main numerical challenges have to be overcome. First, one has to face the exponential scaling of complexity with the number of atoms. In this regards, ML-MCTDHB follows the philosophy of the state-of-the-art method called Multi-Configurational Time-Dependent Hartree method for Bosons (MCTDHB) \cite{Alon2008}, which employs a dynamically optimized single particle basis. Within the class of wave function propagation methods, there are alternatively also the highly successful tensor-network methods, which have been extended to also tackle two-dimensional problems (see e.g. \cite{Orus2014}), but are tailored to discrete systems (see e.g. \cite{Schmidt2007,Delande2014} for applications based on the tight-binding approximation of continuous space).

The second challenge concerns the separation of length scales, which have to be resolved, and are directly related to the question how to appropriately describe short-range interactions between ultracold bosons in three spatial dimensions.
We will review fundamental problems concerning the (regularized) delta interaction approach in numerical many-body methods \cite{Esry1999, Doganov2013, Castin2004}, and show that a carefully chosen Gaussian model potential is most suitable for mimicking the short-range interactions within the ML-MCTDHB approach. The thereby separated length scales require a huge amount of grid points, which would make a corresponding MCTDHB calculation extremely costly or even unfeasible, since the time-dependent three-dimensional single-particle basis states are expanded on a product grid. In contrast to this, ML-MCTDHB makes use of the reduced correlations between the spatial directions in elongated traps \cite{Tacla2011}, which leads to a more efficient representation of the time-dependent single-particle basis. As a consequence, the computational effort w.r.t.\ the number of grid points $q_s$ in the direction $s=1,2,3$ is reduced from $q_1^2 q_2^2 q_3^2$ to $c_1q_1^2+c_2q_2^2+c_3q_3^2$ with problem-dependent prefactors $c_s$. Since ML-MCTDHB reduces to the mean-field theory in a limiting case, our method can also be employed for efficiently solving the Gross-Pitaevskii equation for elongated traps.

For illustrating the power of our method, we study the many-body tunnelling dynamics of bosons in elongated double well traps beyond the mean-field approximation, focusing on the emergence of particle correlations in the crossover from 3D to 1D.
Many works have previously studied bosonic tunnelling in double well set-ups, e.g. \ either theoretically within a mean-field approximation, using two-mode approximations, by rigorous one-dimensional simulations or experimentally (see \cite{Leggett2001,Gati2007} and refs. therein)

We investigate the dynamics of the population imbalance in dependence of the transversal confinement and unravel differences between our converged numerical result and the mean-field predictions as well as beyond mean-field calculations based on an adiabatic separation of the transversal degrees-of-freedom.
In particular, we find that the latter approach cannot resolve the shape of the density distribution even of the initial state, resulting in a different subsequent dynamics.

This work is structured as followed: To be self-contained and provide the proper embedding, we first review the challenges of and approaches to the description of short-range interactions within many-body methods in three spatial dimensions (section \ref{section_2}). The approach of our choice, an appropriate finite-range Gaussian model potential, is discussed in detail and we present scattering calculations in order to relate the parameters of the model potential to the physically relevant s-wave scattering length. In section \ref{section_3}, we address the ML-MCTDHB method and explain how it benefits from the lack of strong correlations between the spatial directions in elongated traps. 
Thereafter, we apply in section \ref{section_4} ML-MCTDHB to a tunnelling problem with a detailed discussion of the convergence and unravel three-dimensional as well as beyond mean-field effects. Finally, we conclude and give an outlook in section \ref{section_5}.


\section{Short-range interaction of bosonic ensembles in 3D: challenges and approaches} \label{section_2}

When simulating the correlated quantum dynamics of short-range 
interacting ultracold bosonic atoms in elongated three-dimensional traps, two 
main challenges have to be faced: (i) the exponential scaling of complexity 
with the particle number and (ii) the separation of different length scales, which 
have to be resolved. While the former challenge is addressed in section 
\ref{section_3}, we deal with the issue of how to treat short-range
interactions here. First, we elaborate on the following fundamental problem: when using numerical methods 
operating in the laboratory frame and being based on a finite product basis, one 
cannot accurately model the interaction by a (regularized) delta potential in 
order to remove the shortest length scale of the problem at hand. Thereafter, we 
review different approaches to model short-range interactions and argue that an 
appropriately normalized Gaussian model potential is well suited for our 
ML-MCTDHB method.
Using such a finite-range model potential, however, implies that the different separated length scales 
have to be resolved, and in section \ref{section_3}
we will explain how ML-MCTDHB copes with this issue for elongated traps in an efficient 
manner.

\subsection{Length scales} \label{subsection_2A}

In the following, we consider an elongated dilute ensemble of $N$ ultracold 
bosonic atoms of mass $m$, which is governed by the time-dependent 
Schr\"odinger equation
\begin{align} \label{eq:SchroedingerEquation}
i \hbar \partial_t 
\Psi(\mathbf{r}_1,\mathbf{r}_2,...,\mathbf{r}_N,t) 
&= {H}\Psi( \mathbf{r}_1,\mathbf{r}_2,...\mathbf{r}_N,t) \nonumber \\
&= \left[\sum_{i=1}^{N} {H}_{0}^{(i)}+\sum_{1\leq i<j\leq N}  {W}^{(ij)}\right]  \Psi(\mathbf{r}_1,\mathbf{r}_2,...,\mathbf{r}_N,t).
\end{align}
Here, $\mathbf{r}_i=\left(x_i,\ y_i,\ z_i\right)^{T}$ refers to the position of the $i$-th atom in Cartesian coordinates, $H_{0}^{(i)} = -\frac{\hbar^{2}}{2m} \nabla_{\mathbf{r}_i}^{2} + V(\mathbf{r}_i)$ denotes its single-particle Hamiltonian and ${W}^{(ij)}=W(\mathbf{r}_i-\mathbf{r}_j)$ is the interaction potential of the $i$-th with the $j$-th atom, $j\neq i$.
In order to model an elongated trap, a harmonic confinement is assumed $V(\mathbf{r}_i) = \frac{1}{2}m\omega_{\bot}^{2} \left(  x_i^2 + y_i^{2} \right) + \frac{1}{2}m\omega_{\parallel}^{2} z_i^{2}$. 
The longitudinal (transversal) trap frequency is $\omega_{\parallel}$ ($\omega_{\bot}$) corresponding to a characteristic length scale $l_{\parallel}=\sqrt{\hbar/(m\omega_{\parallel})}$ ($l_{\bot}=\sqrt{\hbar/(m\omega_{\bot})}$ ) and the aspect ratio is defined as $\eta=\omega_{\bot}/\omega_{\parallel}$.
Experimentally \cite{Berrada2016,Albiez2005a} the aspect ratio can vary from one  to several hundred, where $\eta=1$ describes an isotropic confinement and $\eta\gg1$ a highly elongated trap.
Thus, elongated traps are characterized by two characteristic length scales, with  $l_{\bot}<l_{\parallel}$.
We note that our following considerations are not restricted to  harmonic potentials.

The interaction between two neutral bosons ground state bosonic atoms is given by a Van-der-Waals potential, whose range is much smaller than the typical interparticle distance in dilute ultracold systems \cite{Pethick2008}.
As we will see, removing this smallest length scale by introducing delta interaction potentials with the same s-wave scattering properties in the far field \cite{Fermi1936,Huang1957} suffers in higher dimensions from principle problems within many if not most computational methods \cite{Esry1999}.
So we have to mimic the desired scattering behaviour by an appropriate model potential of finite range $\Sigma$ leading to three length scales, which have to be resolved naturally: $\Sigma\ll l_{\bot}<l_{\parallel}$, covering also the interparticle distance.

\subsection{Modelling short-range interactions} \label{subsection_2B}

\subsubsection{Bare delta interaction}
The simplest way to remove the interaction range from the description of our system is to invoke the bare delta function, $W(\mathbf{r})=g\delta(\mathbf{r})$, where $\mathbf{r}=\mathbf{r}_1-\mathbf{r}_2$ denotes the relative position of the two colliding particles and the interaction strength $g$ is chosen in such a way that $W(\mathbf{r})$ features the desired s-wave scattering length \cite{Pethick2008,Pitaevskii2003}.
Following the lines of \cite{Castin2001}, we review why this is a valid approach in one spatial dimension, which, however, breaks down in three dimensions, where the plane delta function does not scatter \cite{Esry1999,Doganov2013,Castin2001}.

In general, the solution of the stationary two-body scattering problem in the far field is given by an incoming plain wave plus an outgoing spherical wave multiplied with the scattering amplitude $f$
\begin{equation} \label{eq:ScatteringSolution}
\Psi(\mathbf{r})=e^{i\mathbf{kr}}+f(k,\mathbf{r}/r)\,\frac{e^{ikr}}{r}
\end{equation}
where $r=|\mathbf{r}|$  and $\mathbf{r}$ is the relative coordinate vector as well as $E=\hbar^{2}k^{2} / 2 \mu$ denotes the energy of the incoming wave with the reduced mass $\mu$ and wave vector $\mathbf{k}$ of modulus $k$.
For the bare delta interaction, the scattering amplitude can be calculated via the T-matrix formalism \cite{Castin2001}, and is given by $f(k,\mathbf{r}/r)=-\frac{\mu}{2\pi\hbar^2} \frac{g}{1-gJ(E+i0+)}=f(k)$ with 
\begin{equation} \label{eq:Integral}
  J(E+i0+)=\lim_{\epsilon\rightarrow0+}\int\frac{d^{D}k'}{(2\pi)^{D}}\frac{1}{E+i\epsilon-\frac{\hbar^{2}k'^{2}}{2\mu}}
\end{equation}
where $D$ denotes the dimension of the scattering problem.
For ultracold collisions,
the scattering amplitude can be well approximated by its zero-energy limit, the s-wave scattering length
\begin{equation} \label{eq:a0}
a  \equiv -\lim_{k \rightarrow 0 } f(k)
\end{equation}
For $D=1$, the integral \eqref{eq:Integral} is convergent and the bare delta interaction can be used without any limitations in analytical and numerical calculations.
In three dimensions, however, the real part of this integral \eqref{eq:Integral} is ultra-violet divergent and results in $a=0$, such that no scattering takes place in three dimension.

In order to appropriately describe scattering in three dimensions in terms of zero range-potentials, two different strategies can be followed: (i) using the regularized delta interaction $W(\mathbf{r})=g\delta(\mathbf{r})\partial_r r$ \cite{Fermi1936,Huang1957}, instead of the bare delta interaction, which avoids the ultra-violet divergence in \eqref{eq:Integral} or (ii) regularizing the integral \eqref{eq:Integral} for the bare delta interaction by an ultra-violet cut-off and appropriately renormalizing its interaction strength $g$ depending on this cut-off.

\subsubsection{Regularized delta interaction}
Considering strategy (i) first, the regularized  delta interaction has been introduced to incorporate appropriate boundary conditions for the scattering process in three dimensions \cite{Fermi1936,Huang1957,Derevianko2005, Olshanii2001}.
In particular, the $1/r$ divergence of the scattering solution (\ref{eq:ScatteringSolution}) is taken properly into account by the regularization operator $\partial_r r$, which ensures a finite scattering amplitude $f(k) = -\frac{a}{1+ika}$ with $a=\frac{\mu}{2\pi\hbar^2}g$ as the corresponding s-wave scattering length \eqref{eq:a0} \cite{Castin2001}.
Furthermore, in \cite{Farrell2010,Stock2005} it has been shown that the regularized delta interaction emerges, if one considers a proper  zero-range limit of a three-dimensional finite square well or a delta-shell potential, respectively.

The regularized delta interaction is of great use for analytical calculations 
as demonstrated by the solution of the bosonic two-body problem in isotropic 
\cite{Busch1998} and anisotropic \cite{Idziaszek2006} harmonic traps. For most 
numerical many-body methods, however, there is a fundamental problem, which 
stems from the fact that they are formulated in second quantization, and  
operate in the laboratory frame. Thereby, these methods rely on a
finite single particle basis $\chi_k(\mathbf{r}_i)$, $k=1,...,M$ and thus 
effectively on a finite (possibly symmetrized) Hartree product basis as one can 
see e.g.\ from the matrix elements of the 
interaction operator $\langle\chi_i\chi_j|W|\chi_q\chi_p\rangle$. By 
transforming the derivative $\partial_r$ into the laboratory frame, one can easily show 
that 
\begin{align}
g \delta(\mathbf{r}) \partial_r \left( r \chi_i(\mathbf{r}_1) \chi_j(\mathbf{r}_2) \right) =
g \delta(\mathbf{r}) \Big[ 
& \chi_i(\mathbf{r}_1) \chi_j(\mathbf{r}_2) + \\\nonumber
&+\frac{\mathbf{r}_1-\mathbf{r}_2}{2}
\left( \nabla_{\mathbf{r}_1} - \nabla_{\mathbf{r}_2} \right)
\chi_i( \mathbf{r}_1) \chi_j(\mathbf{r}_2)  \Big],
\end{align}
where $\mathbf{r}=\mathbf{r}_1-\mathbf{r}_2$.
Therefore, $ g\delta(\mathbf{r})\partial_r \left(r \chi_i(\mathbf{r}_1) \chi_j(\mathbf{r}_2) \right) 
=g\delta(\mathbf{r})\chi_i(\mathbf{r}_1)\chi_j(\mathbf{r}_2)$, which is 
a direct consequence of the fact that a single 
Hartree product cannot feature a $1/r$ singularity and so the action of the 
regularization operator $\partial_r(r\,\cdot)$ becomes trivial. Moreover, since
$\partial_r(r\,\cdot)$ commutes with any finite sum, the action of 
the regularized delta potential on any available two-body state 
$\sum_{i,j=1}^M a_{ij} \chi_i(\mathbf{r}_1)\chi_j(\mathbf{r}_2)$, 
$a_{ij}\in\mathbb{C}$, is equivalent to the action of the bare delta 
potential, which has important consequences:

(i) Using the regularized delta potential is equivalent to the bare one for laboratory frame methods being based 
on a finite single-particle basis.
So if one knows that the total state can be 
approximated well by the Gross-Pitaevskii mean-field ansatz, i.e.\ a single 
Hartree product, the regularized delta interaction can be safely replaced by 
the bare one, bearing in mind that correlations on short length scales are not 
accurately described for finite $N$ \cite{Erdos2007}.

(ii) If one, however, expects non-trivial correlations or 
is not sure about the applicability of a mean-field ansatz, the regularized 
delta potential does not help since one converges to the non-interacting 
solution when increasing the basis size \cite{Doganov2013}.

(iii) One can only benefit from the correct scattering properties of the 
regularized delta potential if one analytically or numerically utilizes a 
correlated two-body basis. This idea has been implemented in the form of an 
effective interaction potential, being constructed from the exact solution of 
the two-body problem (see e.g.\ \cite{Christensson2009,Lindgren2014} and 
references therein), which provides good predictions for the eigenenergies of 
the many-boson problem. Since we, however, are interested in quantum dynamics 
and operate with a dynamically optimized single-particle basis (see section 
\ref{section_3}), evaluating the matrix elements of the effective interaction 
potential by a six-dimensional integration at each instant in time would lead to large and definitely infeasible computational costs.

\subsubsection{Renormalization of the bare delta interaction}
The second strategy (ii) is to regularize the integral \eqref{eq:Integral} and to renormalize the bare coupling strength such that the regularization is compensated and the correct scattering physics is reproduced: This can be achieved by e.g.\ a discretization of coordinate space \cite{Castin2004}, i.e.\ introduction of a high-momentum cut-off (see also e.g.\ \cite{Cavalcanti1999,Mitra1998,Zinner2012,Rontani2008}).
The main intrinsic truncation procedure within of our ML-MCTDHB method, however, is not the discretization of space but considering only a reduced number of dynamically optimized single-particle basis states (see section \ref{section_3}).
As a consequence one would have to evaluate the renormalization condition, involving e.g.\ the scattering $T$-matrix in the truncated basis, at every instant in time, which would not only be costly but also conceptually difficult, since ML-MCTDHB is based on a truncated single-particle basis in the laboratory frame while the $T$-matrix relates to the relative coordinate frame. Therefore, a similar renormalization could only be established in 1D for strong interactions in special cases \cite{Ernst2011}.

\subsubsection{Finite-range model potentials}
As a consequence of the discussion above, we have to replace the interatomic interactions by a finite-range model potential satisfying the following requirements:
(i) the potential must be short-range, i.e.\ its range $\Sigma$ must constitute the smallest length scale in the problem at hand;
(ii) its s-wave scattering length must be easily tunable via the model potential parameters;
(iii) within the considered numerical many-body method, the model potential should lead to the lowest possible computational costs.

Due to requirement (i),  the trap does not influence the scattering processes, which may then be 
regarded as if taking place in free space \cite{Tiesinga2000,Block2002,Pade2003}.
Moreover, the shape of the model potential does not matter in this case \cite{Blume2002a} and, thus, a variety of different candidates can be conceived, including model potentials with both attractive and repulsive parts.
 
Regarding requirement (ii), (partially) attractive model potentials which feature bound state(s) are in principle easily tunable and cover the full range of possible scattering lengths, since scattering resonances can be exploited by shifting a bound state near the threshold \cite{Taylor2000}.
Bound states, however, in general imply strong interparticle correlations in the laboratory frame, which are difficult to handle, i.e. converge numerically.
To avoid these correlations and a strongly energy-dependent scattering length, we exclusively restrict ourselves to repulsive model potentials in the following.

Regarding requirement (iii), the evaluation of the two-body interaction matrix elements $\langle \chi_{i}\chi_{j}|W|\chi_{k}\chi_{l}\rangle$ in general is computationally very costly, since a six-dimensional spatial integration has to 
be performed, requiring $q_{1}^{2} q_{2}^{2} q_{3}^{2}$ operations, with $q_s$ denoting the number of grid points in the $s$-th dimension.
To reduce the computational effort three strategies are possible:
First, one can approximate the model potential $W$ by a sum over w.r.t.\ the interacting particles separable operators, namely by $W_p(\mathbf{r}_1-\mathbf{r}_2)=\sum_{i=1}^p c_{i} w^{(1)}_i(\mathbf{r}_1) w^{(2)}_i(\mathbf{r}_2)$, using a Schmidt decomposition \cite{Schmidt1907,Jackle1996,Jackle1998}.
For $p=q_1q_2q_3$, this decomposition becomes exact, i.e.\ $W=W_p$. If now $W_p$ approximates $W$ well for much less terms, i.e.\ $p\ll q_1q_2q_3$, the two-body interaction matrix costs only $2p$ integrals over three spatial dimensions, i.e.\ $2p\,q_1q_2 q_3$.
Empirically, however, we found that a large number of terms $p$ is required for a fair representation of short-range potentials, otherwise $W_p$ exhibits unphysical oscillations even at large distances
\footnote{
These oscillations can be damped manually by considering the average $\propto\sum_{i=1}^p W_i$ as the model potential
}, 
resulting in long-range interactions and possibly even bound states.
Therefore, this strategy is not further followed.

In the second strategy, one makes use of the fact that $W$ depends only on 
the relative coordinate of the interacting particles and not on their 
centre of mass position. A succession of Fast Fourier Transformations then 
allows to reduce the number of operations from $q_{1}^{2} q_{2}^{2} q_{3}^{2}$ 
to $\propto q_1q_2q_3\log(q_1q_2q_3)$ \cite{Sakmann2011} (see also appendix 
\ref{Appendix_IMEST}).

The third strategy, finally, is adapted to situations in which the three-dimensional single-particle states $|\chi_j\rangle$ are expanded w.r.t.\  one-dimensional basis states $|\phi^{(s)}_j\rangle$ for the different spatial directions $s$ (as it is done efficiently in ML-MCTDHB, see section \ref{section_3}).
 Similarly to the first strategy, one can then (approximately) unravel $W(\mathbf{r}_1-\mathbf{r}_2)$ into a sum of 
operators that are separable w.r.t.\ spatial directions, namely $\sum_{j_1=1}^{p_1}\sum_{j_2=1}^{p_2}\sum_{j_3=1}^{p_3} c_{j_1j_2j_3}\, W^{(1)}_{j_1}(x_1-x_2) \, W^{(2)}_{j_2}(y_1-y_2) \,W^{(3)}_{j_3}(z_1-z_2)$, which can be achieved by the POTFIT algorithm \cite{Jackle1996,Jackle1998}.
In this way, the six-dimensional integration in $\langle \chi_{i}\chi_{j}|W|\chi_{k}\chi_{l}\rangle$ becomes a sum over two-dimensional integrations $\langle\phi^{(s)}_q|W^{(s)}_{j_s}|\phi^{(s)}_p\rangle$ and the efficiency of this scheme depends on $p_i$ as well as on how many one-dimensional basis states $|\phi^{(s)}_j\rangle$ are needed for convergence.

Since our ML-MCTDHB method aims at an efficient representation of $|\chi_i\rangle$ in terms of the $|\phi^{(s)}_j\rangle$, we follow this third strategy in combination with the second strategy for evaluating $\langle\phi^{(s)}_q|W^{(s)}_{j_s}|\phi^{(s)}_p\rangle$.
Moreover, we restrict ourselves to model potentials $W(\mathbf{r}_1-\mathbf{r}_2) = W^{(1)}(x_1-x_2) W^{(2)}(y_1-y_2) W^{(3)}(z_1-z_2)$, which are separable in Cartesian coordinates, to further reduce the number of summations.
Demanding isotropy, a natural choice for $W$ is the 
Gaussian interaction potential in three dimensions $W_{G}(\mathbf{r}_{1}-\mathbf{r}_{2}) = h e^{-\left(\mathbf{r}_{1}-\mathbf{r}_{2}\right)^{2}/\sigma^{2}}$, with height $h$ and width $\sigma$.
The interaction strength can be adjusted by $h$, however, increasing the height of the Gaussian potential increases also its range $\Sigma(h,\sigma)$ and thus we may violate the relation $\Sigma\ll l_{\bot}$ for strong interactions.
Therefore, we normalize the width of the Gaussian w.r.t. a small energy scale of our system $\epsilon$ by demanding $W_{RG}(\mathbf{r})=\epsilon$ if $|\mathbf{r}|=\sigma$ , such that $\Sigma$ is independent of the height $h$:
\begin{equation}    \label{eq:GaussianInteraction}
W_{RG}(\mathbf{r}_{1}-\mathbf{r}_{2})=h e^{-\ln( 
h/\epsilon)\,\frac{\left(\mathbf{r}_{1}-\mathbf{r}_{2}\right)^{2}}{\sigma^{
2}}}.
\end{equation}
In the inset of figure \ref{fig:ScatteringLengths}, we compare the two Gaussian profiles, $W_{G}$ and $W_{RG}$ with the 
interaction potential $W_T(\mathbf{r})=g\Theta(\sigma-\left| \mathbf{r} \right|)$, which is not separable.
Increasing the height, the renormalized Gaussian $W_{RG}$ approaches the shape of the theta function $W_{T}$, whose width is limited by $\sigma=\Sigma$. In contrast to this, the range of $W_{G}$ grows unlimited with increasing $h$ such that the assumptions of both free space scattering and short-range interactions will be violated. Thus, we favour $W_{RG}$.

\subsection{Tunability of model interaction potentials} \label{subsection_2C}

Given the s-wave scattering length $a$ of a physical scenario, which, in principle, is experimentally tunable over a broad range via Feshbach resonances \cite{Chin2010}, we have to choose the free parameters of our model 
potential $W$ such that the latter constitutes a short-range interaction potential of the same $a$.
Since, only repulsive potentials are considered, $a$ is naturally bounded by $\Sigma$  from above, which inevitably restricts the realizable scattering lengths to $a\ll l_\perp$.
We numerically determine the dependence of the energy-dependent phase-shift $\delta(k)$ on the model potential parameters by solving the s-wave scattering problem in free space in relative coordinates with the Numerov method \cite{Numerov1927} and comparing with the non-scattered solution in the far field.
Then, we extrapolate the $k \rightarrow 0$ behaviour of $\delta(k)$ in order to determine 
$a$ \cite{Taylor2000}.

\begin{figure}
\centering
\includegraphics[width=\linewidth]{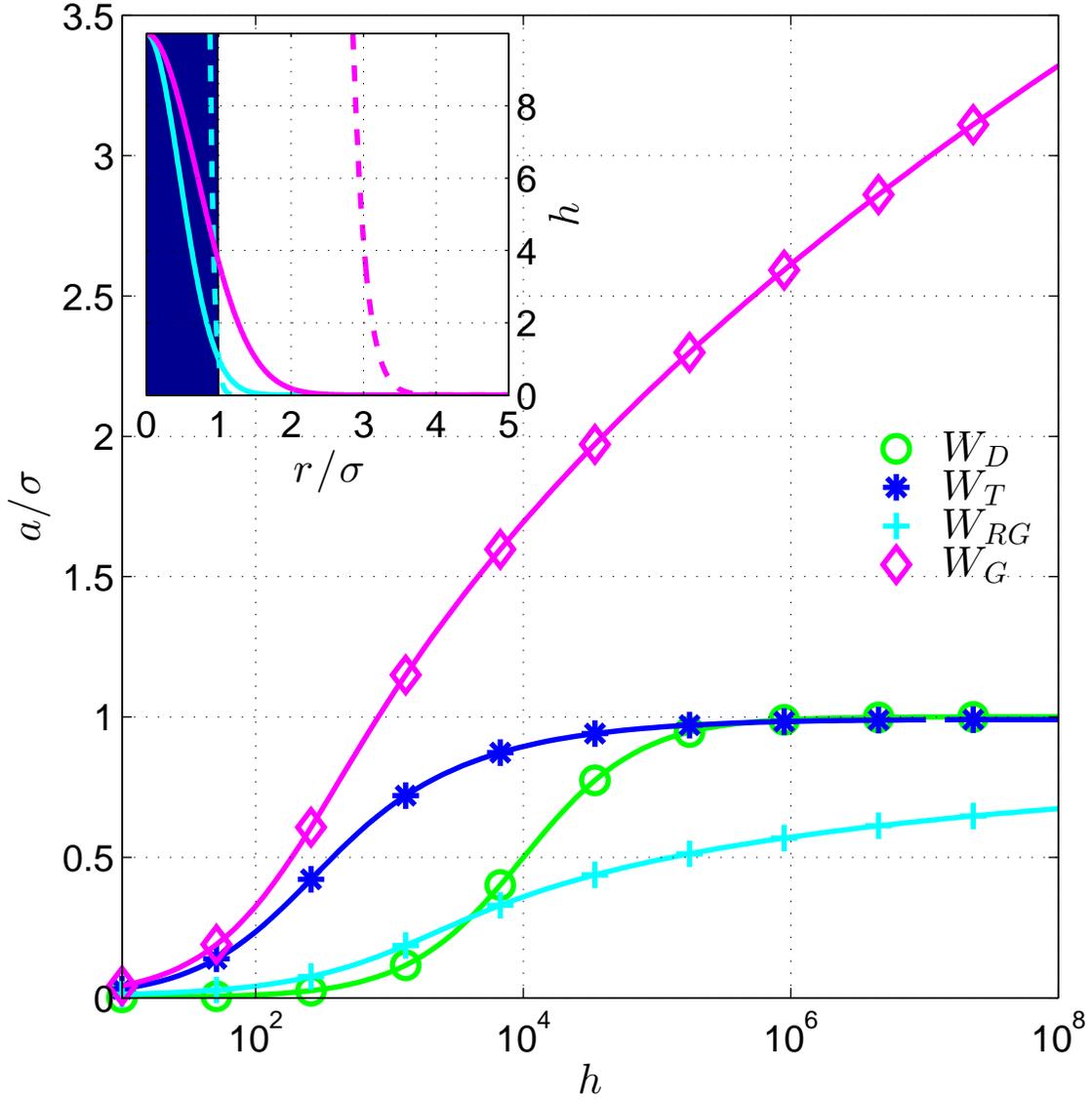}
\caption{ \label{fig:ScatteringLengths}
(Colour online)
Scattering lengths for different interaction potentials:
green circles $W_{D}(\mathbf{r})=h\delta(\sigma-|\mathbf{r}|)$,
blue stars $W_{T}(\mathbf{r})=h\Theta(\sigma-|\mathbf{r}|)$,
cyan plus $W_{RG}(\mathbf{r})=h\,\exp\left(-\sqrt{\ln(h/\epsilon)}\,\mathbf{r}^{2}/\sigma^{2}\right)$ with $\epsilon=1$,
and magenta diamonds $W_{G}(\mathbf{r})=h\exp\left(-\mathbf{r}^{2}/\sigma^{2}\right)$.
Please note the logarithmic scale of the x-axis.
The scattering length is scaled in units of the characteristic interaction length scale $\sigma$.
The inset  shows the shape of these interaction potentials.
The blue shaded area refers to the theta interaction potential and serves as a reference.
The cyan lines show $W_{RG}$ and the magenta lines $W_{G}$ for $h=10$ (solid line) and  $2 \cdot10^{3}$ (dashed line), respectively.
}
\end{figure}

For the different interaction potentials $W_T$, $W_G$, $W_{RG}$  and the delta shell model $W_D(\mathbf{r})=g\delta(\sigma-\left| \mathbf{r} \right|)$, we show the dependence of the s-wave scattering length on the potential height in figure \ref{fig:ScatteringLengths}.
As analytically expected \cite{Taylor2000}, $a$ saturates for 
$W_{T}$ and $W_D$, such that  the three-dimensional scattering length becomes equal to the range of the interaction potential, $a=\sigma=\Sigma$.
Whereas the scattering length of $W_G$ (as well as its range) diverges with increasing $h$, the scattering length of $W_{RG}$ converges slowly towards the limit $a=\sigma=\Sigma$, as expected.
In doing so, the interaction range does not interfere with other length scales in the system.
Although the achievable scattering lengths are limited to small values, we 
choose $W_{RG}$ as our model potential because of the beneficial computational 
properties discussed above. This implies that strong interparticle 
correlations can be achieved only by increasing the particle density of the 
bosonic ensemble $n$ such that $a n^{1/3}\gtrsim 1$ or by providing 
sufficiently high excitation energies in the initial state for triggering 
dynamical quantum depletion.
However, already in the weakly interacting regime 
intriguing beyond mean-field effects can be found \cite{Streltsov2013,Klaiman2014,Streltsova2014a,Weiner2014,Tsatsos2015,Klaiman2016}.

\section{Computational approach to many-body quantum dynamics} \label{section_3}

The second numerical challenge is the exponential scaling of complexity w.r.t. the particle number, when solving the many-body Schr\"odinger equation.
We face this challenge with the so-called Multi-Layer Multi-Configurational 
Time-Dependent Hartree method for Bosons (ML-MCTDHB) \cite{Cao2013, Kronke2013}.
This ab-initio method becomes numerically exact for large enough basis sets, 
and can treat both bosonic single-component ensembles and mixtures of different 
bosonic species in one or more spatial dimensions.
In this method, the many-body wave function is expanded w.r.t. 
variationally optimized time-dependent many-body basis states, which span the 
relevant subspace of the Hilbert space at each 
instance in time.
How it faces the exponential scaling of the number of configurations 
for single component systems lies at the heart of the Multi-Configurational 
Time-Dependent Hartree method for Bosons (MCTDHB) \cite{Alon2008}, which 
is shortly reviewed in section \ref{subsection_3A} (see  \cite{Beck2000, 
Meyer2011,Meyer2009} for a detailed review of this class of wave function-propagation methods).
Thereafter, we explain how ML-MCTDHB makes use of the separation of energy 
scales in different spatial directions, as it occurs in elongated traps for 
instance, in order to obtain a better scaling w.r.t. the number of grid 
points (section \ref{subsection_3B}).
Indeed, ML-MCTDHB takes advantage of 
the absence of strong correlations between the different spatial directions in 
such a case.
This aspect of ML-MCTDHB was briefly addressed in \cite{Cao2013} but major developments were required here to overcome the resulting numerical challenges.
In the last section \ref{subsection_3C}, we show how ML-MCTDHB can be used 
for the non-trivial task of solving the three-dimensional GPE for elongated traps.

\subsection{The MCTDHB method and higher dimensions} \label{subsection_3A}

In MCTDHB \cite{Alon2008}, the many-body wave function for $N$ bosons is expanded w.r.t.\ a set of time-dependent permanents, i.e bosonic number states 
\begin{equation} \label{eq:Expantion_TL}
|\Psi(t)\rangle=\sum_{\vec{n}|N}A_{\vec{n}}(t)|\vec{n}\rangle_{t}
\end{equation}
These permanents are labelled by an integer vector $\vec{n}=(n_{1},\ ...,\ n_{i},\ ...,\ n_{M})$, where $n_{i}$ is the occupation number of the \textit{$i$}-th three-dimensional, time-dependent single-particle function (3D-SPF), $\left|\chi_{i}(t)\right\rangle $, which is variationally optimized at each instant in time.
The number $M$ of considered 3D-SPFs  constitutes a numerical control parameter of this method.
The symbol $\left.\vec{n}\right|N$ denotes the summation over all $N$-body permanents.
Then, the 3D-SPF are represented w.r.t.\ a time-independent basis $|U_J \rangle$, for example given by a discrete variable representation (DVR) \cite{Light2000,Beck2000} or a Fast-Fourier Transformation based grid \cite{Kosloff1983,Tannor2007},
\begin{equation} \label{eq:Expansion_PL}
\left|\chi_{j}(t)\right\rangle
=\sum_{J=1}^{Q}\tilde{B}_{jJ}(t)|U_{J}\rangle
\end{equation}
where $\tilde{B}_{jJ}(t)$ denote the time-dependent expansion coefficients and $Q=q_1 q_2 q_3$ refers to the total number of grid points.
The equations of motion for the time-dependent coefficients can be derived, 
using the Dirac-Frenkel variational principle \cite{Dirac1930,Frenkel1932}, 
$\left\langle \delta\Psi\right|i\partial_{t}-{H}\left|\Psi\right\rangle =0$, 
under the constraint $\langle \chi_j | i\partial_{t} |\chi_k \rangle =0$, which ensures the orthonormality of the 3D-SPFs.
Performing the variation w.r.t. the time-dependent coefficients in \eqref{eq:Expantion_TL} and \eqref{eq:Expansion_PL}, leads to \cite{Alon2008}
\begin{equation} \label{eq:EOM_TL}
i \hbar \partial_t A_{\vec{n}}=\sum_{\vec{m}|N}\left\langle 
\vec{n}\right|{H}\left|\vec{m}\right\rangle A_{\vec{m}}
\end{equation}
\begin{equation} \label{eq:EOM_PL}
i \hbar\partial_t \tilde{B}_{iI}
=\sum_{k,l=1}^{M}\sum_{L=1}^{Q}
\left\langle 
U_{I}\right|\left[\left(1-{P}^{(\chi)}\right)\left({H}_{0}^{(1)}+\left[\rho^
{(\chi)}\right]_{ik}^{-1}\langle{W}\rangle_{kl}^{(\chi)}\right) \right] 
\left|U_{L}\right\rangle \tilde{B}_{lL}
\end{equation}
Here, to shorten the notation, we drop the time-dependence for simplicity.
The projector, ${P}^{(\chi)}=\sum_{i=1}^{M}\left|\chi_{i}\rangle\!
\langle \chi_{i}\right|$, projects onto the co-moving subspace spanned by the instantaneous 3D-SPFs.
According to equation (\ref{eq:EOM_PL}), the 3D-SPF can only rotate into the orthogonal complement of the subspace which they are spanning. 
The dynamics of the 3D-SPFs is driven by both the one-body Hamiltonian $H_0^{(1)}$ and the interactions among the atoms, which are represented by the second term
in the bracket on the right-hand-side of (\ref{eq:EOM_PL}). Here, $\rho_{ik}^{(\chi)}$
denotes the one-body density matrix\footnote{\label{fnote_dmat}We note that in the (ML-)MCTDHB terminology, the one-/ and two-body density matrix are proportional to the transposed of the physical reduced one-/ and two-body density matrix} in the 3D-SPF representation, which can be calculated by $\rho_{ik}=\langle\Psi|{a}_{i}^{\dagger}{a}_{k}|\Psi\rangle$/N, 
where ${a}_{i}^{\dagger}$ $\left({a}_{i}\right)$ creates (destroys) a 
boson in the state $\left|\chi_{i}(t)\right\rangle$. 
The inverse of the one-body density matrix causes weakly occupied single particle functions to rotate faster than strongly occupied ones \cite{Beck2000}.
Finally, $\langle{W}\rangle_{ik}^{(\chi)}$ refers to the so-called mean-field operator matrix, which can be written as 
$\langle{W}\rangle_{ik}^{(\chi)}=\sum_{j,l=1}^M\rho_{ijkl}^{(2)}\sum_{I,K=1}^QW_{IjKl}^{
(\chi)}|U_{I}\rangle\!\langle U_{K}|$, with
$W_{IjKl}=\sum_{JL}\tilde{B}_{jJ}^*\tilde{B}_{lL}\langle 
U_{I}U_{J}|{W}|U_{K}U_{L}\rangle$ and the two particle density matrix 
$\rho_{ijkl}^{(2)}=\langle\Psi|{a}_{i}^{\dagger}{a}_{j}^{\dagger}
{a}_{k}{a}_{l}|\Psi\rangle/N$.
Thus, the interaction couples different 3D-SPF, weighted by the corresponding two body density 
matrix elements.

The time-independent basis is typically chosen such that the two-body interaction matrix elements $\langle 
U_{I}U_{J}|{W}|U_{K}U_{L}\rangle$ can be easily evaluated analytically or numerically at the beginning of a simulation.
Yet we emphasize that one needs  $(q_1q_2q_3)^2$ operations to calculate the elements $W_{IjKl}$, for each evaluation of the r.h.s. of (\ref{eq:EOM_PL}).
The computational effort can be reduced to $q_1q_2q_3\log(q_1q_2q_3)$ by making use of fast Fourier transformations (see section \ref{subsection_2B} and appendix \ref{Appendix_IMEST}), but the evaluation of these matrix elements then still constitutes the numerical bottle neck for simulations requiring large grids.

If there are as many 3D-SPF as grid points, $M=\prod_{s=1}^3 q_s$, the full CI limit is recovered, where the Hilbert space is only truncated by the discretization of the coordinate space.
Whereas, if only one 3D-SPF is supplied, MCTDHB reduces to the Gross-Pitaevskii mean-field theory
(see also section \ref{subsection_3C}).
A MCTDHB calculation may be regarded as numerically converged, if in particular the expectation values of the observables of interest become  to a certain quantitative degree insensitive to a further increase of the number of basis states, characterized by the numerical configuration $(M|q_1q_2q_3)$. \\

\subsection{ML-MCTDHB for elongated traps} \label{subsection_3B}

The key idea of our ML-MCTDHB approach is to find a more efficient representation of the MCTDHB 3D-SPFs in order to reduce the significant computational costs for the evaluation of the mean-field operator matrix $\langle{W}\rangle_{ik}^{(\chi)}$, which have restricted MCTDHB to small grids so far. This venture is in particular crucial for addressing the 1D-to-3D crossover in elongated traps due to the three separated length scales which have to be resolved (see section \ref{subsection_2A}). Yet exactly in this regime one may anticipate that the transversal and longitudinal degrees-of-freedom are not too strongly correlated since the transversal excitation energies separate from the longitudinal ones.
This motivates us to expand the 3D-SPFs $\left|\chi_{i}(t)\right\rangle $ w.r.t.\ a product of three one-dimensional, time-dependent single particle functions (1D-SPFs), $|\Phi_{\mathbf{J}}(t)\rangle=\bigotimes_{s=1}^3|\phi_{j_{s}}^{(s)}(t)\rangle$, where we have introduced the multi-index $\mathbf{J}= \left( j_1, j_2, j_3 \right)$ with $j_s=1,...,m_s$ and $m_s\leq q_s$
\begin{equation} \label{eq:Expansion_PL2}
\left|\chi_{i}(t)\right\rangle
=\sum_{j_1=1}^{m_1} \sum_{j_2=1}^{m_2} \sum_{j_3=1}^{m_3}  B_{ij_1j_2j_3}(t)
\bigotimes_{s=1}^3|\phi_{j_{s}}^{(s)}(t)\rangle
\equiv\sum_{\mathbf{J}}B_{i\mathbf{J}}(t)|\Phi_{\mathbf{J}}(t)\rangle
\end{equation}
The 1D-SPFs in turn are expanded w.r.t. a time-independent, one-dimensional basis $|u^{(s)}_r\rangle$, $r=1,...,q_s$, i.e. represented on a one-dimensional grid
\begin{equation} \label{eq:temp3}
|\phi_{j}^{(s)}(t)\rangle=\sum_{r=1}^{q_s} C_{jr}^{(s)}(t)|u^{(s)}_r\rangle.
\end{equation}
Thus we have introduced an additional truncation: If the correlations between the spatial directions are not too strong, the 3D-SPFs can be represented well by taking into account only $m_s\ll q_s$ 1D-SPFs, as we will also show exemplarily in section \ref{section_4}. This is in particular advantageous for elongated traps where due to the separation of energy scales one may take less 1D-SPFs for the transversal directions than for the longitudinal direction into account, i.e.\ $m_1,m_2<m_3$.
In an even more extreme case when we fix $m_1=m_2=1$, the transversal degrees-of-freedom adiabatically separate
This will be a good approximation for sufficiently large aspect ratios $\eta$, where beyond mean-field effects are strongly dominated by the population of various longitudinal 1D-SPFs. But also for modest $\eta$, it is promising to compare fully converged ML-MCTDHB calculations with simulations in which $m_1=m_2=1$ is fixed and $m_3$ is increased until convergence. In this way, one compares full 3D simulations with quasi-1D ones and can identify the impact of correlations between the spatial directions.

Using the Dirac-Frenkel variational principle, one finds that the $A_{\vec{n}}$ coefficients still obey (\ref{eq:EOM_TL}), while the dynamics of the 3D-SPFs expansion coefficients is now governed by
\begin{equation} \label{eq:EOM_PL_ML_B}
i \hbar\partial_t {B}_{i\mathbf{I}}
=\sum_{k,l=1}^{M}\sum_{\mathbf{L}}
\left\langle 
\Phi_{\mathbf{I}}\right|\left[\left(1-{P}^{(\chi)}\right)\left({H}_{0}^{(1)}+\left[\rho^
{(\chi)}\right]_{ik}^{-1}\langle{W}\rangle_{kl}^{(\chi)}\right) \right] 
\left|\Phi_{\mathbf{L}}\right\rangle {B}_{l\mathbf{L}}
\end{equation}
additional equations of motion can been derived for the coefficients $C_{ir}^{(s)}$. 
\begin{equation}  \label{eq:EOM_SPL}
i\hbar {\partial_t}C_{ir}^{(s)} = 
\sum_{r'=1}^{q_s}
\sum_{j,k=1}^{m_s}
\langle u^{(s)}_r|(1-{P}^{(s)})\left({h}_{0}^{(s)}+
\left[\rho^{(s)}\right]_{ik}^{-1}\,\left[\langle{\bar V}\rangle_{kj}^{(s)}+
\langle{W}\rangle_{kj}^{(s)}\right]\right)|u^{(s)}_{r'}
\rangle C_{jr'}^{(s)}
\end{equation}
Here, ${P}^{(s)}$ denotes the projector $\sum_{i=1}^{m_s}|\phi_i^{(s)}\rangle\!\langle\phi_i^{(s)}|$ and $\rho_{ik}^{(s)}$ refers to the density matrix of the $s$-th degree-of-freedom of a single bosons (see appendix \ref{Appendix_Ingredients}, eq. \eqref{appA_rho} for an explicit definition). The operator $h_0^{(s)}$ contains all terms of $H_0^{(1)}$ which act non-trivially on the $s$-th coordinate but as a unit operator on the $s'$-th one, $s'\neq s$. The terms of $H_0^{(1)}$ which couple the $s$-th direction to the other ones enter equation (\ref{eq:EOM_SPL}) via the mean-field operator matrix $\langle{\bar V}\rangle_{kj}^{(s)}$, while the interaction among the other atoms induces the mean-field operator matrix $\langle{W}\rangle_{kj}^{(s)}$ (see appendix \ref{Appendix_Ingredients}).

Regarding the interactions, it is important to realize that summations over the grid do only enter the calculation of the interaction induced mean-field operator matrix $\langle{W}\rangle_{kj}^{(s)}$. By means of the algorithm reviewed in appendix \ref{Appendix_IMEST}, this results in costs scaling like $q_s\log q_s$ for separable potentials, where the prefactors strongly depend on the numbers of 1D-SPFs [see equation (\ref{eq_mf_op_mat_spl})]. However, once the basic ingredients for $\langle{W}\rangle_{kj}^{(s)}$ are known, the mean-field operator matrix $\langle{W}\rangle_{kl}^{(\chi)}$ can directly be calculated without further summations over the grid [see equation (\ref{eq_mf_op_mat_pl})], which is in stark contrast to the corresponding calculation in MCTDHB. 

Our ML-MCTDHB method formally reduces to MCTDHB if as many 1D-SPFs are supplied as there are grid points $m_i=q_i$. As a result,
the maximal strength of correlations between the spatial directions, which can be resolved, is solely limited by the grid. 
But as we have argued before and as we will see exemplarily in section \ref{section_4}, the correlations between the spatial directions are rather weak in many relevant situations. Under such circumstances, ML-MCTDHB will be more efficient than MCTDHB and allows in particular for employing much larger grids. Comparing the scaling of ML-MCTDHB and MCTDHB w.r.t.\ the number of grid points is involved and depends on the details of the implementation. However, one may at least state that MCTDHB requires $M\,q_1q_2q_3$ coefficients for representing the 3D-SPFs, while $M\,m_1m_2m_3+\sum_{s=1}^3m_sq_s$ coefficients are needed in ML-MCTDHB. So if one can achieve convergence for sufficiently small $m_s\ll q_s$, ML-MCTDHB will be much more efficient.

Finally, there is a word of caution in order here: In contrast to the MCTDHB theory, the ML-MCTDHB equations of motion do not automatically conserve symmetries involving transformations of two or three coordinates, such as rotations or reflections, but only if the simulation is converged w.r.t.\ $(m_1,m_2,m_3)$  \cite{Cao2013}. This disadvantage can be cured by choosing symmetry-adapted coordinates, of course. In such coordinates, however, the considered interaction model potential $W$ will in general not be separable. For reducing the computational effort, one thus has to apply the POTFIT algorithm w.r.t. the ``spatial directions'' to $W$ \cite{Jackle1996,Jackle1998} (see the discussion in section \ref{subsection_2B}) or to find an appropriate model potential, which is separable in the new coordinates if feasible at all.

\subsection{ML-MCTDHB as an efficient solver for the 3D Gross-Pitaevskii 
equation} \label{subsection_3C}
The equations of motion (\ref{eq:EOM_PL}) can be reduced to the GPE \cite{Gross1961,Pitaevskii1961}, if only one 3D-SPF  ($M=1$) is supplied and if one is converged w.r.t. $(m_1,m_2,m_3)$. 
According to the considerations in section \ref{subsection_2B}, the short-range interaction can then be accurately modelled by the bare delta potential ${W}=g\delta(\mathbf{r}_{1}-\mathbf{r}_{2})$ as it is usually done within mean-field theory.
Computationally, the bare delta potential is very advantageous for the ML-MCTDHB wave function ansatz (\ref{eq:Expansion_PL2}) since it is separable w.r.t.\ the spatial dimensions (in Cartesian coordinates) and since the integral over one degree-of-freedom can be carried out analytically.
In this way, approximately $\sum_{s=1}^3 m_s^4q_s$ summations regarding the grid plus additional summations over the states 
$|\phi^{(s')}_i\rangle$ with $s' \ne s$ (see equation \eqref{eq_mf_op_mat_spl}) 
are required for calculating the mean-field operator matrix of $\langle{W}\rangle_{kj}^{(s)}$.
Thus, both the representation of the condensate wave function and the evaluation of its equations of motion scale linearly w.r.t. the number of grid points with $m_s$-dependent prefactors\footnote{Non-local operators such as the kinetic energy result in a $q_s^2$ scaling if one does not employ an Fast Fourier Transformation based grid.}. In contrast to this, solving the Gross-Pitaevskii equation on a product grid, as it usually done, results in a $q_1q_2q_3$ scaling.
Consequently, our method presented in section \ref{subsection_3B} can be used to solve the 
GPE very efficiently if the coupling between the spatial dimensions is not too strong, i.e.\ $m_s\ll q_s$, as it is the case for elongated traps (see \cite{Tacla2011}).


\section{Double well tunnelling}\label{section_4}

In this section, we analyse the ground state and the dynamics for the dimensional cross over from 3D to 1D of a bosonic ensemble tunnelling in an elongated double well.
We are interested in the emergence of particle correlations with varying  aspect ratio.

In section \ref{subsection_4A}, we describe the physical set-up.
Then, the initial ground state is analysed in section \ref{subsection_4B}, with a particular emphasis on the convergence of the simulations, followed by a discussion of the density and the population imbalance between the right and left well.
In section \ref{subsection_4C}, we explore the corresponding tunnelling dynamics with a focus on particle correlations and the temporal evolution of the density.
We do not aim at an exhaustive study of the cross-over from 1D to 3D with varying parameters but want to demonstrate the numerical feasibility of three-dimensional simulations of bosonic ensembles beyond mean-field, with this illustrating example, for our computational ML-MCTDHB approach.

\subsection{Set-up} \label{subsection_4A} 

We study an ensemble of $N=14$ bosons loaded into an elongated double well trap,
which is modelled by the potential $
V(\mathbf{r}_{i})= V_{trap}(\mathbf{r}_{i}) + V_{barrier}(\mathbf{r}_{i}) + V_{tilt}(\mathbf{r}_{i})$.
Employing harmonic oscillator units w.r.t.\ the longitudinal direction, the harmonic trap reads 
$V_{trap}(\mathbf{r}_{i})=1/2 \; \eta^2 (x_i^2 + y_i^2) + 1/2 \; z_i^2$. A Gaussian barrier 
$V_{barrier}(\mathbf{r}_{i})=H\,\exp(-\mathbf{r}_i^2 / S^2)$, experimentally realizable by a focused blue-detuned laser, with fixed height $H=10$ and width $S=0.4$, separates the trap into a left and right half. To obtain the initial ground state with a population imbalance, an additional potential $V_{tilt}(\mathbf{r}_{i})=d_{z}z_i$ with $d_{z}=-0.1$
is taken into account.
Then the ML-MCTDHB equations of motion are propagated in imaginary time to relax to the ground state, which contains a surplus of particles on the right side.
The interaction between the particles is modelled by the normalized Gaussian $W_{RG}$ (see section \ref{subsection_2B}). Choosing the height $h=125$, width $\sigma=0.1$ and energy scale $\epsilon=1$, 
the scattering length\footnote{This scattering length is obtained by solving the free-space scattering problem on 
a grid for the relative coordinate, which is derived from the laboratory  frame grid employed in our ML-MCTDHB calculations.} is fixed to $a=0.0048$ according to section \ref{subsection_2C}.
The width $\sigma$ is the smallest physical length scale in the system, which is resolved numerically by our grid spacing of $0.02$ (in all directions).
Employing a Fast-Fourier-transformation based grid \cite{Kosloff1983}, we take $500$ ($400$) grid points for the longitudinal (for each transversal) direction into account.

\subsection{Ground-state analysis}\label{subsection_4B}

Before investigating the properties of the ground states, we give a detailed analysis of the convergence.

\subsubsection{Convergence study}

In order to check the convergence of the simulations, we vary the number of 3D-SPFs $M$ and 1D-SPFs $(m_{1},m_{2},m_{3})$ for fixed physical parameters, and compare the ML-MCTDHB results for the observables of interest.
A simulation is converged, if the observables of interest do not change upon increasing the number of SPFs further.
One has to check carefully the convergence in ML-MCTDHB, because an interdependence between $M$ and $m_s$ can occur.

For the considered cylindrically symmetrical trap, we may choose the number of transversal 1D-SPFs to be equal: $m_1=m_2$. For nearly isotropic traps, i.e.\ $\eta\gtrsim 1$, one uses $m_1=m_2=m_3$.
If the transversal trap is tightened, less transversal 1D-SPF are needed, and all particle correlations, if existent, are handed over to the population of longitudinal 1D-SPFs, thus a good choice is to set $M=m_3 \ge m_1=m_2$. In the following, each simulation is characterized by the numerical configuration $\mathcal{C}=(M|m_{1}m_{2}m_{3})$.

A first indicator for convergence can be obtained by a spectral analysis of certain reduced density operators, i.e. the eigenvalues (natural populations) $a^{(\chi)}_i$ and $b_i^{(s)}$ of the reduced density operator of a single boson $\rho^{(\chi)}$ and of the $s$-th degree-of-freedom of a single boson $\rho^{(s)}$ and their eigenvectors (natural orbitals).
The smallest natural population $a^{(\chi)}_i$ and $b_i^{(s)}$ may serve as a practical measure for how many 3D-SPF and 1D-SPF are needed, respectively, and the natural-population distribution is sensitive to the presence of particle and spatial correlations, respectively. Such conclusions from the natural orbitals are rigorous for already converged simulations and otherwise only indicative (see \cite{Cosme2015} for a critical discussion).
Within our normalization, we have $0\leq a^{(\chi)}_i,b_i^{(s)}\leq 1$ and $\sum_ia^{(\chi)}_i=\sum_ib^{(s)}_i=1$. Moreover, we label the natural populations in decreasing sequence.

\begin{figure} 
\centering
\includegraphics[width=\linewidth]{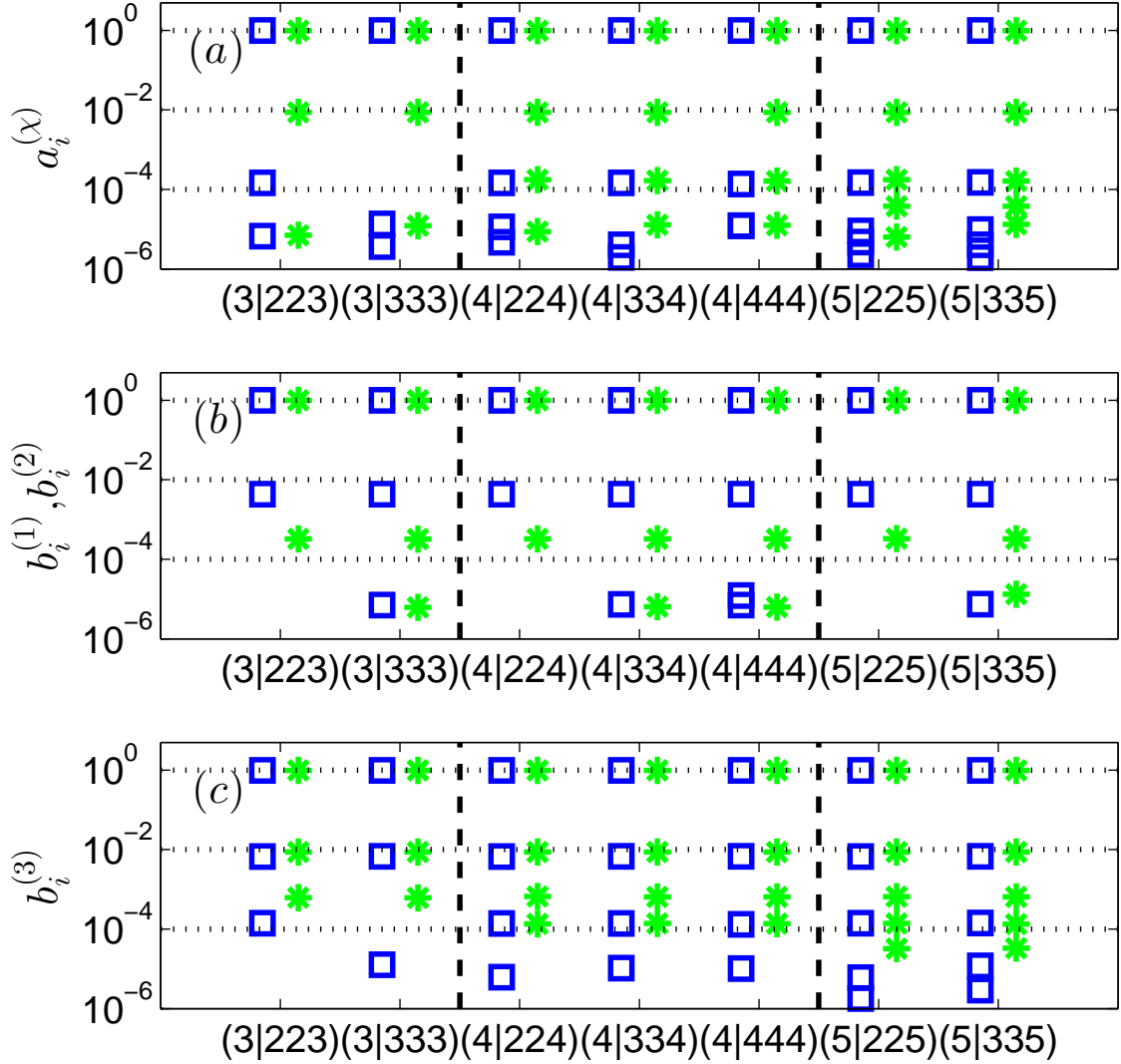}
\caption{(Colour online)
(a) shows the natural populations $a_i^{(\chi)}$ of $\rho^{(\chi)}$, and (b) and (c) presents the natural populations $b_i^{(s)}$ of $\rho^{(s)}$, respectively, for two different aspect ratios $\eta=2$ (blue squares) and $\eta=8$ (green stars).
The superscript $s$ denotes the dimension.
The horizontal axis provides different numerical configurations $\mathcal{C}$, where $M=m_3$ is increased for various $m_1=m_2$ kept fixed.
The black dashed vertical line separates different configurations, where $M=m_3$ is increased by one.
}
\label{fig:GS_NatPop}
\end{figure}

In figure \ref{fig:GS_NatPop}, we show the natural populations $b_{i}^{(s)}$ and $a_{i}^{(\chi)}$ for two aspect ratios $\eta=2$ and $\eta=8$  and different numerical configurations $\mathcal{C}$.
Adding an additional 3D-SPF, the two most dominant natural populations $a^{(\chi)}_{1,2}$ change by only $\sim10^{-4}$, i.e.\ not significantly (see figure \ref{fig:GS_NatPop}a).
A tighter trap depletes $a^{(\chi)}_1$ in favour of $a^{(\chi)}_2$, indicating already emerging particle correlations.
Adding further 1D-SPFs, the two most dominant natural populations of $\rho^{(s)}$, $b^{(s)}_{1,2}$, are not significantly changed and corrections take place of the order of $10^{-4}$.
As expected, the second dominant natural population $b^{(1,2)}_{2}$ of the transversal directions is stronger populated for a more isotropic trap, $\eta=2$, than for $\eta=8$, which implies stronger spatial correlations (see figure \ref{fig:GS_NatPop} b). Finally, figure \ref{fig:GS_NatPop}c shows that the spectrum of $\rho^{(3)}$ is rather robust w.r.t.\ adding more 3D- and 1D-SPFs.

\begin{figure} 
\centering
\includegraphics[width=\linewidth]{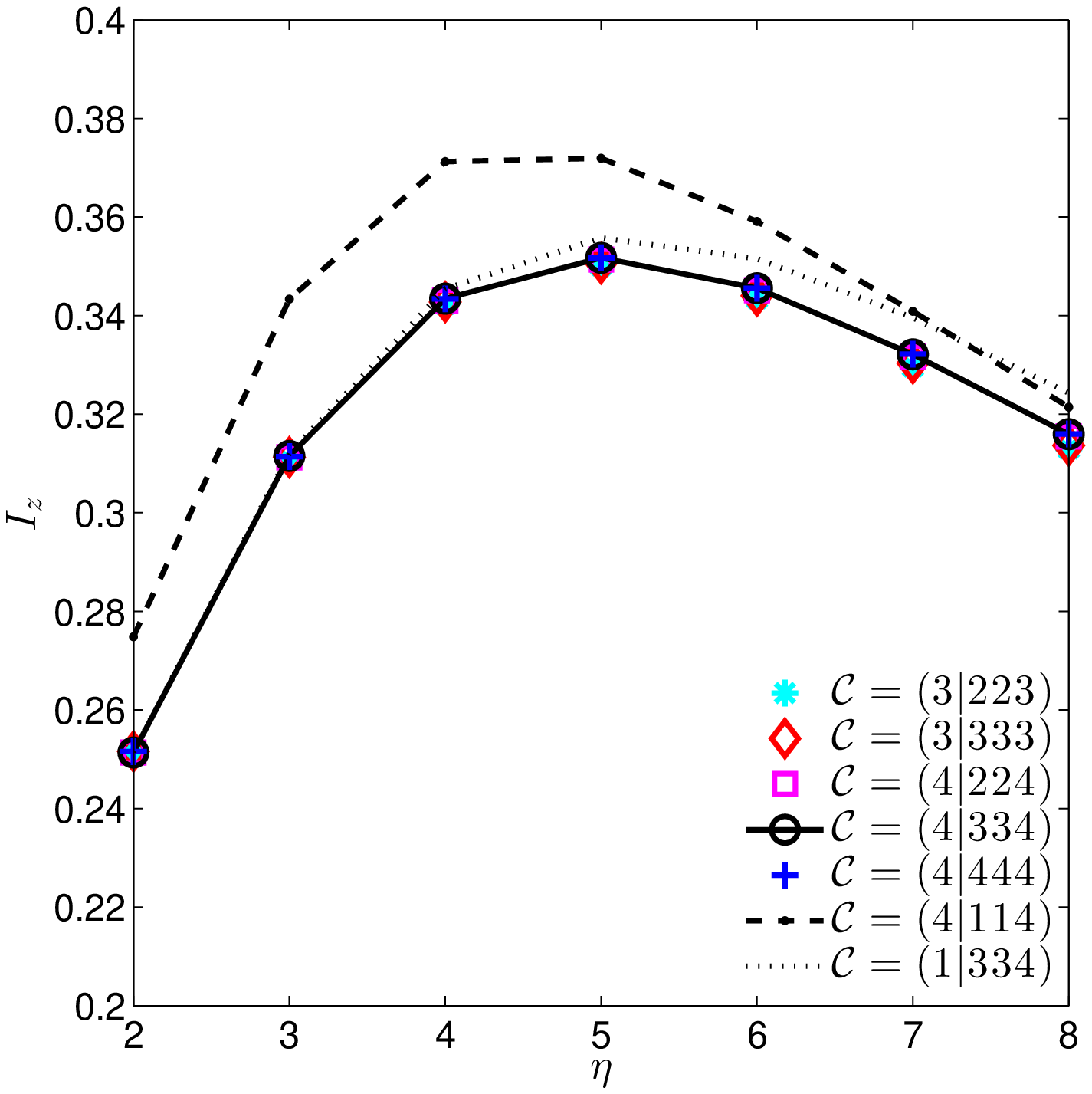}
\caption{(Colour online)
Shown are the population imbalances for different numerical configurations $\mathcal{C}$ with respect of the aspect ratio.
The lines connecting the points are plotted as a guide to the eye.
The black dashed, dotted and solid lines are the mean-field $\mathcal{C}_{MF}=(4|114)$ , quasi 1D $\mathcal{C}_{q1D}={1|334}$ and fully converged $\mathcal{C}_{FC}=(4|334)$ simulations.
}
\label{fig:GS_PopImbalance}
\end{figure}

Second, we study the convergence of our ground-state calculation by comparing the population imbalance between the two wells,  $I_z = (N_L-N_R)/N$, where $N_{L,R}=\sum_{i=1}^N\langle \Theta( \pm z_i) \rangle$, for various configurations $\mathcal{C}$ in dependence on the aspect ratio $\eta$ (figure \ref{fig:GS_PopImbalance}).
We see that nearly all simulations show excellent agreement for the population imbalance,except for the configurations $\mathcal{C}_{MF}=(1|334)$ and $\mathcal{C}_{q1D}=(4|114)$.
The configuration $\mathcal{C}_{FC}=(4|334)$ is regarded as fully converged for this observable.
The configuration $\mathcal{C}_{MF}$ corresponds to a mean-field configuration ($M=1$), where the $m_s$ are increased until convergence, i.e.\ to the ground state of the 3D GPE. One can clearly see that 3D GPE agrees well with the fully converged results for $\eta < 4$, while the mean-field results deviate for more anisotropic traps since interparticle correlations become important. In contrast to this, the configuration $\mathcal{C}_{q1D}$ corresponds to an adiabatic separation of the transversal degrees-of-freedom while resolving interparticle correlations by bringing the simulation to convergence w.r.t. $M=m_3$. This quasi one-dimensional simulation approaches the fully converged results for increasing $\eta$, but even for $\eta=8$, significant deviations remain.

\subsubsection{Configuration dependent ground state properties}

\begin{figure}  
\centering
\includegraphics[width=\linewidth]{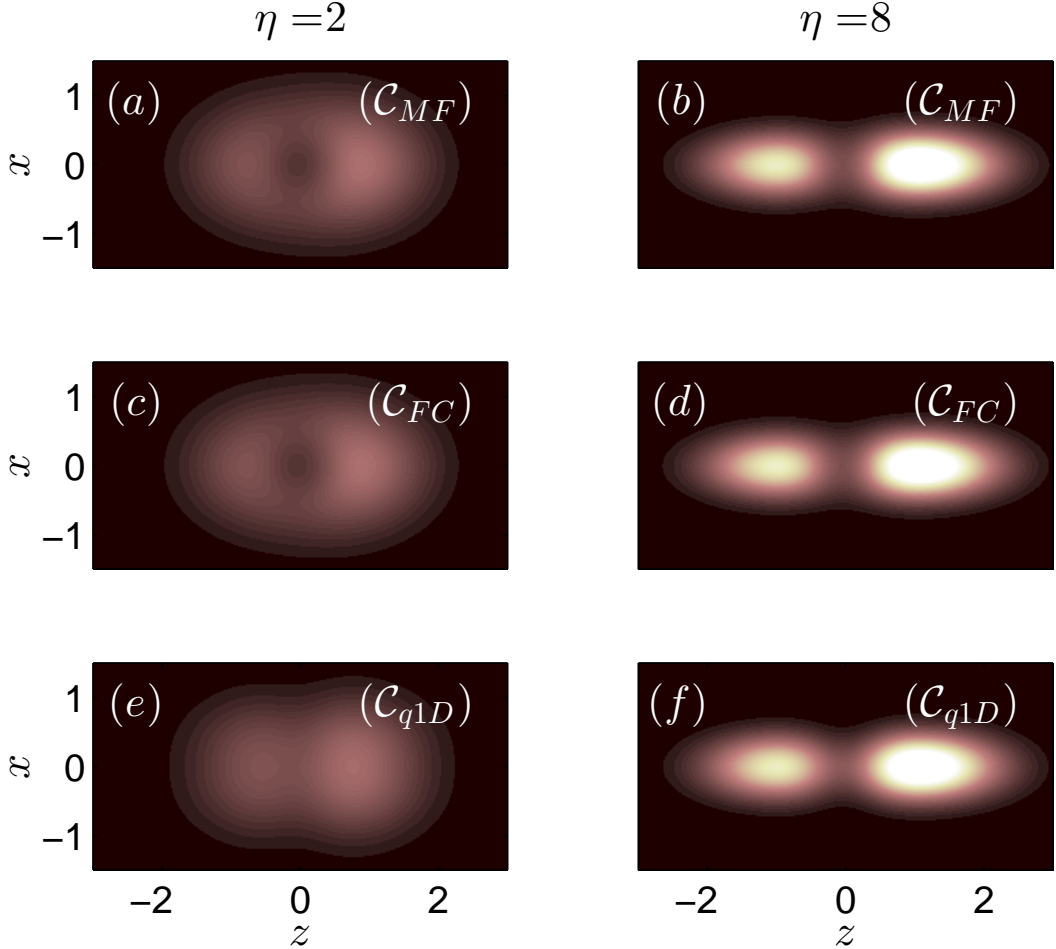}
\caption{(Colour online)
Cut through the density $\rho(x,y=0,z,t)$ for different aspect ratios $\eta=2$ (left column), $\eta=8$ (right column)  and configurations $\mathcal{C}_{MF}$ (first row), $\mathcal{C}_{FC}$ (second row) and $\mathcal{C}_{q1D}$ (third row) w.r.t. the longitudinal $z$ and transversal dimension $x$. 
Harmonic oscillator units are used.
The range of the colour bar is from zero (black) to 0.8 (white).
}
\label{fig:GS_Density}  
\end{figure}

Next, we explore in some detail the validity of the adiabatic separation of the transversal degrees-of-freedom or the double.
Let's consider a cut of the three-dimensional ground state density $\rho(x,y=0,z)$ for the three numerical configurations $\mathcal{C}_{MF}$, $\mathcal{C}_{q1D}$ and $\mathcal{C}_{FC}$.
The mean-field (figure \ref{fig:GS_Density}a) and fully converged (figure \ref{fig:GS_Density}c) density profile agree well for $\eta=2$, whereas they differ significantly in their geometry from the quasi 1D simulation $\mathcal{C}_{q1D}$ (figure \ref{fig:GS_Density}e).
This is interesting, since the second dominant transversal natural orbital is only weakly populated with $b_{2}^{(1)}=b_{2}^{(2)}\sim 4\cdot10^{-3}$ (see figure \ref{fig:GS_NatPop}), indicating less spatial correlations. Nevertheless, these further orbitals are necessary to describe the dip in the density induced by the barrier barrier $V_{barrier}(\mathbf{r}_{i})$.
This density dip cannot be resolved in the quasi 1D simulation due to the crude adiabatic separation, i.e. the single variationally optimized SPF in each transversal direction has a shape independent of the for longitudinal position.

Increasing the transversal trap frequency restricts the wave function stronger in the transversal direction, $\sim 1/(2\eta)$, and if it is smaller than the width of the barrier, the barrier can be Taylor expanded $V_{barrier}(\mathbf{r}_{i}) \sim H \exp(-z_i^2/S^2)$. 
No spatial correlations are induced by this approximated barrier any more, and the adiabatic separation is a good approximation, as can be seen by comparing the density profiles for $\mathcal{C}_{MF}$ (figure \ref{fig:GS_Density}b), $\mathcal{C}_{FC}$ (figure \ref{fig:GS_Density}d) and $\mathcal{C}_{q1D}$ (figure \ref{fig:GS_Density}f) for the aspect ratio $\eta=8$. In this regime, only the interaction could induce spatial correlations, which however is prevented by the transversal excitation gap.

In summary, the differences in the ground state density for weak transversal confinement cause a different initial population imbalance and thus induces a different tunnelling dynamics, as we shall see in the following subsection.

\subsection{Tunnelling dynamics} \label{subsection_4C}

To trigger the tunnelling dynamics, we switch off the tilted potential $V_{tilt}(z)$ at $t=0$ and propagate the many-body wave function in real time, with the intention to study its dynamical features.

\begin{figure}
\centering
\includegraphics[width=\linewidth]{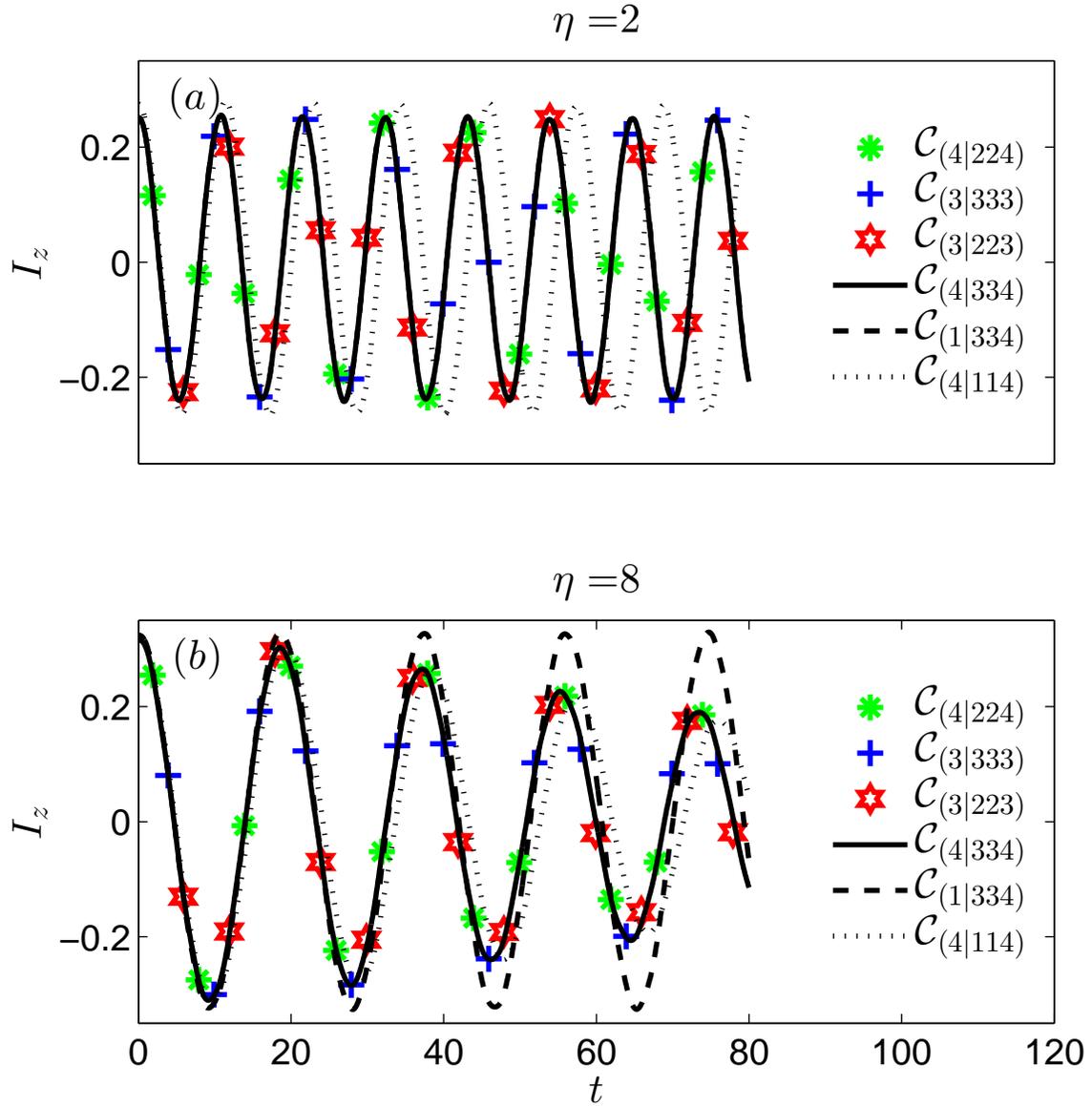}
\caption{  (Colour online).
Temporal evolution of the population imbalance $I_z(t)$  for the two aspect ratios $\eta=2$ (a) and  $\eta=8$ (b).
Shown are the following numerical configurations 
$\mathcal{C}=(4|224)$ (green stars),
$\mathcal{C}=(3|333)$ (blue plus signs), 
$\mathcal{C}=(3|223)$ (red hexagram).
$\mathcal{C}=(4|334)$ (black line),
$\mathcal{C}=(1|334)$ (black dashed line),
$\mathcal{C}=(4|114)$ (black dotted line).
Note that in (a) the black dashed line is covered by the solid black line.
Harmonic oscillator units are used.
}
\label{fig:Dyn_PopImbalance} 
\end{figure}

In order to ensure the convergence of the simulations, let us inspect the time evolution of the population imbalance $I_z(t)$ for different configurations $\mathcal{C}$ in two different trap geometries with aspect ratios $\eta=2$ and $\eta=8$, respectively.
In figure \ref{fig:Dyn_PopImbalance}, the population imbalance shows excellent agreement between the numerical configurations $\mathcal{C}=(4|224)$, $(3|333)$, $(3|223)$ and $(4|334)$ (see figure \ref{fig:Dyn_PopImbalance}).
Hereafter, the numerical configuration $\mathcal{C}=(4|334)$ is referred to the fully converged simulation $\mathcal{C}_{FC}$.

\begin{figure}  
\includegraphics[width=\linewidth]{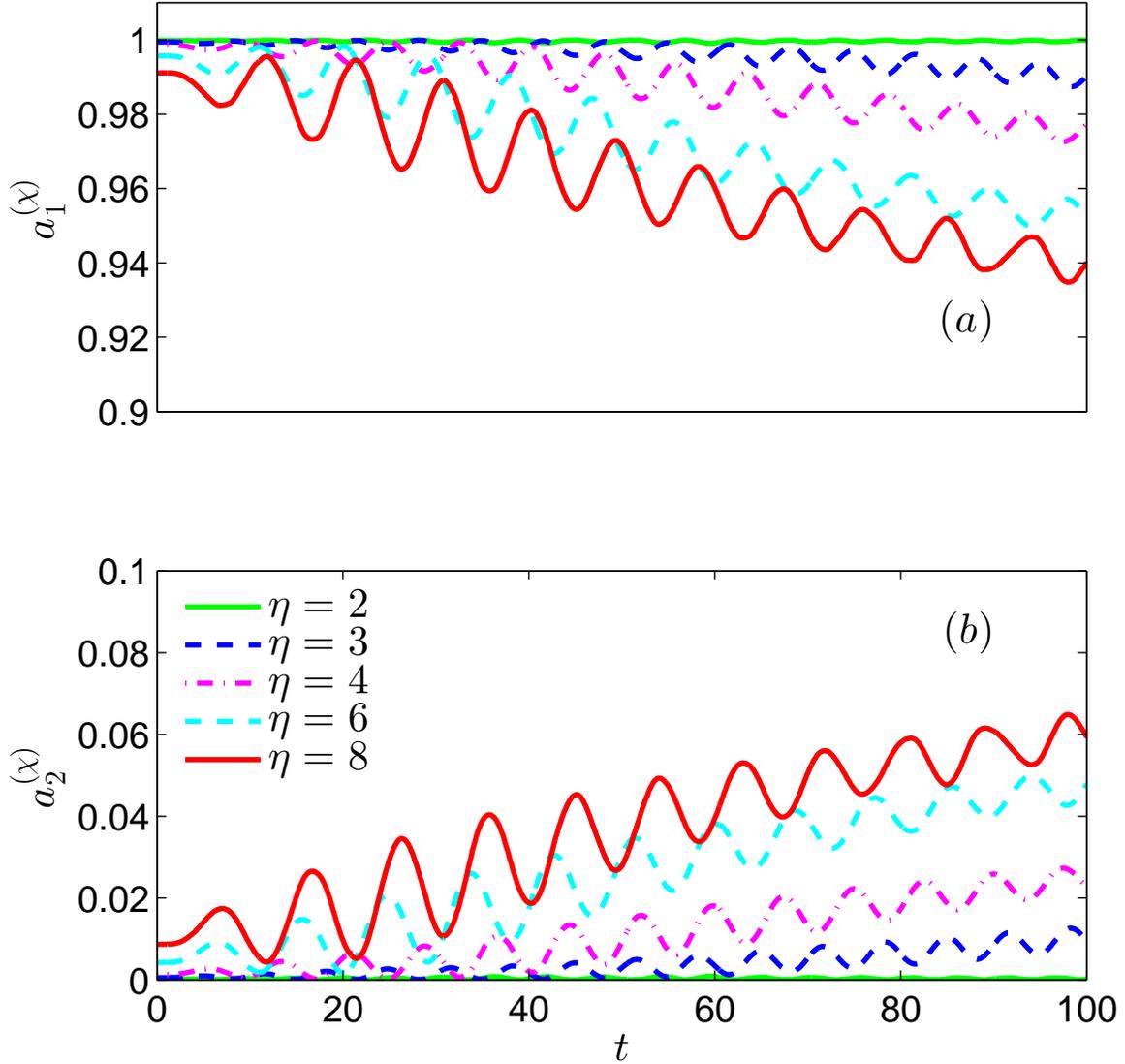}
\caption{(Colour online)
The natural populations $a^{(\chi)}_1$ (a) and $a^{(\chi)}_2$ (b) w.r.t. time for different aspect ratios.
Harmonic oscillator units are used.
Both subfigures have the same colour coding. 
}
\label{fig:Dyn_Natpop}  
\end{figure}

\begin{figure}  
\includegraphics[width=\linewidth]{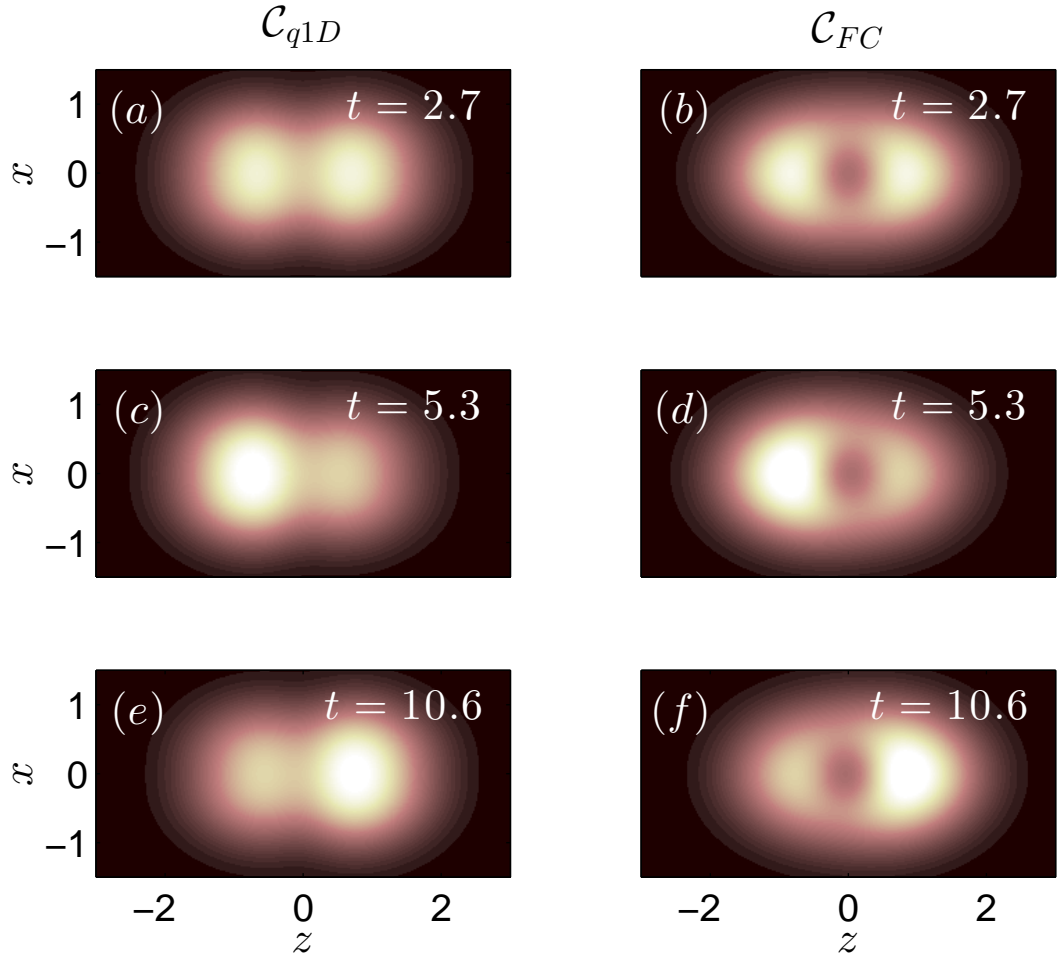}
\caption{(Colour online)
Cut through the density $\rho(x,y=0,z,t)$ for different times, $t=2.9$, $t=5.3$, $t=10.6$  and configurations $\mathcal{C}_{q1D}$ (left column) and $\mathcal{C}_{FC}$ (right column) w.r.t. the longitudinal $z$ and transversal dimension $x$. 
Harmonic oscillator units are used.
The aspect ratios is $\eta=2$.
The range of the colour bar is from zero (black) to 0.2 (white).
}
\label{fig:Dyn_Density} 
\end{figure}

For weak transversal confinement $\eta=2$, the first natural populations of $\mathcal{C}_{FC}$ is very close to one $a_1^{(\chi)} \sim 1$ (see figure \ref{fig:Dyn_Natpop}a), indicating that no particle correlations are present in the system.
Thus, the fully converged simulation can be described within a mean-field approximation, which can also been seen in the almost identical temporal evolution of the  population imbalance of the two configurations $\mathcal{C}_{MF}=(1|334)$ and $\mathcal{C}_{FC}$ (see figure \ref{fig:Dyn_PopImbalance}a). 
In contrast to, the quasi one-dimensional configuration $\mathcal{C}_{q1D}=(4|114)$ fails to describe the right dynamics of the population imbalance for weak transversal confinement, because only one transversal mode cannot resolve correctly the initial density profile (see section \ref{subsection_4B}), which impacts the resulting tunnelling behaviour.
Snapshots of the temporal evolution of the density cut $\rho(x,y=0,z,t)$ for the two different numerical configurations $\mathcal{C}_{q1D}$ (see figures \ref{fig:Dyn_Density}a,d,e) and $\mathcal{C}_{FC}$ (see figure \ref{fig:Dyn_Density}b,d,f) show how the particles tunnel from right $(t=0)$ (see \ref{fig:GS_Density}) to left $(t=5.3)$ and back $(t=10.6)$.

For tighter traps, the second natural population $a_2^{(\chi)}$ increases (see figure \ref{fig:Dyn_Natpop}b), leading to higher particle correlations and thus to beyond mean-field effects.
Therefore, the mean-field configuration fails to describe the system, which possesses damping as well as a different frequency in the population imbalance (see figure \ref{fig:Dyn_PopImbalance}b).
Furthermore, increasing the aspect ratio, the second natural populations $b_2^{(1,2)}$ decrease (not shown), expressing small spatial correlations, while the longitudinal natural population $b_2^{(3)}$ inherit the particle correlations $b_2^{(3)} \sim a_2^{(\chi)}$, indicating a cross over to a quasi one-dimensional configuration.

In summary, increasing the aspect ratio leads to stronger particle and weaker spatial correlations. 
For low aspect ratios, the system can be well-described by the mean-field ansatz, whereas for high aspect ratios (quasi 1D) an effective one-dimensional but beyond mean-field theory is needed.
With our here presented numerical method ML-MCTDHB, we are able to simulate the cover-over from quasi one- to three- dimensional behaviours either within mean-field approximations or taking all particle correlations into account.

\section{Conclusion and outlook}\label{section_5}

We have worked out the ab-initio method ML-MCTDHB \cite{Kronke2013,Cao2013} for the study of the three-dimensional quantum dynamics of bosonic many-body systems out of equilibrium.
In general, this poses two main numerical challenges.
The first one is the exponential scaling of complexity with the number of particles, which can be tackled by the state-of-the-art MCTDHB method.
The second challenge concerns the different length scales in the system, which is on the one hand directly linked to the question, how to describe correctly the short-range atom-atom interaction in three dimensions.
Therefore, we have reviewed the fundamental problems occurring when modelling these interactions via delta potentials and alternative approaches leading us to the usage of finite range potentials.
On the other hand, this challenge includes that an elongated trap induces an additional length scale.
Furthermore, in sufficiently elongated traps, the transversal modes are energetically separated from the longitudinal modes.
Exploiting this feature, we expand the 3D-SPF into a product of one-dimensional time-dependent SPFs, which allows for an efficient description of the different length scales.

For this wave function ansatz, the equations of motion have been derived for the 1D-SPF using a variational principle, ensuring an optimized 1D-SPF basis set at each instant of time.
Convergence is realized by successively increasing the number of 1D-SPFs and 3D-SPFs, which serve as a numerical control.
In the limit of a single 3D-SPF, our method constitutes an efficient way to solve the 3D Gross-Pitaveskii equation for elongated traps.

To illustrate our method, we have analysed an ensemble of bosons tunnelling in a double well for different transversal trap frequencies, thereby studying the cross over from the three- to the quasi one-dimensional case.
This is a regime whose quantum dynamics are notoriously difficult to describe by any method.
We show the necessity of at least two transversal modes, in order to get the density profile and the correct tunnelling behaviour.

For an aspect ratio $\eta=2$, we observe significant spatial correlations while only small particle correlations are present.
As a consequence, this regime can be well described by a mean-field ansatz.
Following, we have monitored the dynamical emergence of particle correlations in the cross over from 3D to 1D accompanied by a reduction of spatial correlations.
In this respect, the quasi 1D regime can be well described by an effective one-dimensional beyond mean-field equation.
In summary, our presented numerically method is able to cover the cross over from quasi one to three dimensions both in the mean-field approximation and for numerical exact simulations. 

A natural next step to take is to explore situations, in which the transversal modes take part in the dynamical evolution, such that we can investigate the transfer of correlations between the dimensions.

\section*{Acknowledgements}
We thank Johannes Schurer, Lushuai Cao  and Hans-Dieter Meyer for many fruitful discussions and Ofir E. Alon for pointing out reference \cite{Sakmann2011}.
P.S. gratefully acknowledges financial support by the Deutsche Forschungsgemeinschaft in the framework of the individual grant Schm 885/20-1.

\newpage
\appendix

\section{Ingredients of the ML-MCTDHB equations of motion} \label{Appendix_Ingredients}

Here, the more involved ingredients for the equations of motion \eqref{eq:EOM_SPL} are explicated.
The density matrix of the $s$-th degree-of-freedom of a single bosons is given by
\begin{equation} \label{appA_rho}
 \left[\rho^{(s)}\right]_{ik} = \sum_{q,p}\rho_{qp}^{(\chi)}\sum_{\mathbf{Q}^{s},\mathbf{P}^{s}}
B_{q\mathbf{Q}^{s}_{i}}^{*}
B_{p\mathbf{P}^{s}_{k}}
\end{equation}
where the sum over e.g.\ $\mathbf{Q}^{s}$ abbreviates a summation over $q_{s'}$ with $s'\neq s$ and $\mathbf{Q}^{s}_{i}$ equals $(q_1,q_2,q_3)$ with $q_s$ replaced by $i$.

Now, we explicate the mean-field operator matrix $\langle{\bar V}\rangle_{kj}^{(s)}$, which is induced by the terms of the one-body Hamiltonian $H_0^{(1)}$, which couple the direction $s$ to $s'\neq s$ (e.g.\ the Gaussian barrier potential in section \ref{section_4}). If we abbreviate these coupling terms with $\bar V^{(s)}$, we find
\begin{equation}
 \langle{\bar V}\rangle_{ik}^{(s)} = \sum_{q,p}\rho_{qp}^{(\chi)}\sum_{\mathbf{Q}^{s},\mathbf{P}^{s}}
B_{q\mathbf{Q}^{s}_{i}}^{*}
B_{p\mathbf{P}^{s}_{k}}\,\langle\Phi_{\mathbf{Q}^{s}}|\bar V^{(s)} |\Phi_{\mathbf{P}^{s}}\rangle
\end{equation}
with e.g.\ $ |\Phi_{\mathbf{P}^{s}}\rangle\equiv\bigotimes_{s'\neq s}|\phi^{(s')}_{p_{s'}}\rangle$. So $\langle\Phi_{\mathbf{Q}^{s}}|\bar V^{(s)} |\Phi_{\mathbf{P}^{s}}\rangle$ is an operator acting on the $s$-th degree-of-freedom.

Given that the interaction potential is separable, $W(\mathbf{r}_1-\mathbf{r}_1) = W^{(1)}(x_1-x_2) W^{(2)}(y_1-y_2) W^{(3)}(z_1-z_2)$,
the interaction induced mean-field operator matrix of the $s$-th degree-of-freedom reads
\begin{equation}\label{eq_mf_op_mat_spl}
\langle{W}\rangle_{ik}^{(s)} = 
\sum_{j,l,q,p} \rho_{qjpl}^{(2)} \sum_{\mathbf{J},\mathbf{L},\mathbf{Q}^{s},\mathbf{P}^{s}}
B_{q\mathbf{Q}^{s}_{i}}^{*}B_{j\mathbf{J}}^{*}
B_{p\mathbf{P}^{s}_{k}}B_{l\mathbf{L}}
\left[\prod_{s'\neq s} W^{(s')}_{q_{s'}j_{s'}p_{s'}l_{s'}}
\right]\langle\phi^{(s)}_{j_{s}}|W^{(s)}|\phi^{(s)}_{l_{s}}\rangle,
\end{equation}
where $W^{(s')}_{q_{s'}j_{s'}p_{s'}l_{s'}}= \langle \phi^{(s')}_{q_{s'}} \phi^{(s')}_{j_{s'}}|W^{(s')}|
\phi^{(s')}_{p_{s'}} \phi^{(s')}_{l_{s'}}\rangle$ and $\langle\phi^{(s)}_{j_{s}}|W^{(s)}|\phi^{(s)}_{l_{s}}\rangle$
constitute a single particle operator. These two basic ingredients can be evaluated very efficiently by the algorithm discussed in appendix \ref{Appendix_IMEST}. Finally, these ingredients can be used to calculate the mean-field operator matrix for the equation of motion of the $B_{i\mathbf{I}}$ equations of motion
\begin{equation}\label{eq_mf_op_mat_pl}
 \langle{W}\rangle_{ik}^{(\chi)}=\sum_{j,l}\rho_{ijkl}^{(2)}\sum_{\mathbf{I},\mathbf{J},\mathbf{K},\mathbf{L}}
B_{j\mathbf{J}}^{*}B_{l\mathbf{L}} 
\prod_{s} W^{(s)}_{i_sj_sk_sl_s}\,
|\Phi_{\mathbf{I}}\rangle\,\langle\Phi_{\mathbf{K}}|.
\end{equation}

\section{Interaction Matrix Evaluation by Successive Transforms} \label{Appendix_IMEST}

In this appendix, we want to present an efficient algorithm to calculate the interaction matrix elements
$W^{(s)}_{ijkl}=\langle\phi_{i}^{(s)}\phi_{j}^{(s)}| W^{(s)} |\phi_{k}^{(s)}\phi_{l}^{(s)} \rangle$ for each dimension $s=1,2,3$.
These interaction matrix elements are needed to construct the mean-field operator matrices [see eq. \eqref{eq_mf_op_mat_spl} and \eqref{eq_mf_op_mat_pl}] for the equation of motions \eqref{eq:EOM_PL_ML_B} and \eqref{eq:EOM_SPL}.
For diagonal interaction potentials, such as used in section 
\ref{subsection_2B}, the computational effort is reduced to $\propto q_s^{2}$ and 
the interaction matrix reads
\begin{equation} \label{eq:IMEST_W}
W_{ijkl}^{(s)} = \int dx \int dx' \ \left( \phi_{i}^{(s)}(x) \right)^{*} \left( \phi_{j}^{(s)}(x') \right)^{*}  \phi^{(s)}_{k}(x)  \phi^{(s)}_{l}(x')  W^{(s)}(x-x')
\end{equation}
The idea, how to reduce the numerical effort of the calculations of the interaction matrix effectively, is adopted from \cite{Sakmann2011}.
Here, we follow their lines of argument.
Using the property that $W^{(s)}$ depends only on $\xi=x-x'$, we can reduce the computational effort by using Fast Fourier Transformations (FFT), which scale with the number of grid points as $\propto q_s\log(q_s)$.
The forward and backward Fourier Transformations of the interaction potential read
\begin{equation} \label{eq:IMEST_FFT1}
W^{(s)}(\xi)  =W^{(s)}(x-x') = \frac{1}{\sqrt{2\pi}}\int dk\; e^{-ik(x-x')} w^{(s)}(k)
\end{equation}
\begin{equation}  \label{eq:IMEST_FFT2}
w^{(s)}(k)=\frac{1}{\sqrt{2\pi}}\int d\xi\; e^{ik\xi}W^{(s)}(\xi)
\end{equation}
Inserting \eqref{eq:IMEST_FFT1} into \eqref{eq:IMEST_W} leads to
\begin{equation}
W^{(s)}_{ijkl} = \frac{1}{\sqrt{2\pi}} \int dx \ \left( \phi_{i}^{(s)}(x) \right)^{*} \phi_{k}^{(s)}(x)
\left[ \int dk \; w^{(s)}(k)
\left( \int dx' \ \left( \phi_{j}^{(s)}(x') \right)^{*} \phi^{(s)}_{l}(x')  e^{-ikx'}  \right)
e^{ikx} \right] \nonumber
\end{equation}
The calculation of this integral can be divided into three steps.
The first step is a FFT of the $x'$ particle coordinate 
$g^{(s)}(k) = \frac{1}{\sqrt{2\pi}} \int dx' \left( \phi_{j}^{(s)}(x')  \right)^{*}  \phi^{(s)}_{l}(x')e^{-ikx'}$, followed by an inverse FFT $G^{(s)}(x) = \frac{1}{\sqrt{2\pi}} \int dk\;  w^{(s)}(k) g^{(s)}(k) e^{ikx}$.
In the last and third step, one has to perform a one-dimensional integration over the $x$ particle coordinate
$W^{(s)}_{ijkl} = \sqrt{2\pi} \int dx \left( \phi_{i}^{(s)}(x) \right)^{*} \phi^{(s)}_{k}(x) G^{(s)}(x)$.
In total, the computational effort is $2\cdot q_s\log q_s+q_s$ for every time step, instead of the previous scaling with $\propto q_s^{2}$.


\bibliography{/afs/physnet.uni-hamburg.de/users/zoq_t/vbolsing/Documents/02_Papers/library.bib}


\end{document}